# Perturbation Series in Light-Cone Diagrams of Green Function of String Field

*Am-Gil Li*[1], *Tae-Song Kim*[1*], *Chol-Man Li*[2], *Song-Jin Im*[3]

[1] *Department of Energy Science, Kim Il Sung University, Pyongyang, DPR Korea*

[2] *E-Library, Kim Il Sung University, Pyongyang, DPR Korea*

[3] *Department of Physics, Kim Il Sung University, Pyongyang, DPR Korea*

**Abstract**

In this paper, we proved the correspondence between Feynman's diagrams in space-time and light-cone diagrams in world-sheet by using only path integral representation on free Green function in the first quantization theory. We also obtained general representation on perturbation series of light-cone diagrams describing split and join of strings.

## 1. Introduction

In 1981, A.M. Polyakov introduced the path integral approach to string theory, using a sum-over-surfaces viewpoint that maintains manifest Lorentz, reparametrization, and conformal invariance [1, 2]. However, we considered functional field as function field of infinitely several variables and generalized the traditional path integral formalism in function field into the novel path integral formalism in functional field, such that we obtained generating functional in string field theory [3].

Numerous calculations are required to obtain the result on concrete process by using Feynman's diagrams on Green function of one string represented by generating functional [3] of string field, but it is difficult practically.

Otherwise, it is also difficult to classify Feynman's diagrams in phase, for a great number of very different diagrams appear. But all these complexities can be overcome if it is converted into light-cone diagrams [4].

The bases that Feynman's diagrams in space-time can be changed into light-cone diagrams in $\sigma-\tau$ world-sheet are as follows:

First it is that free propagating function of string field theory can be obtained in two methods, i.e. in the second quantization theory it is [5]

$$G_0^{(2)}[1,\ 2] = {}_x\langle 0|T(\Phi[x_1]\Phi^*[x_2])|0\rangle_x$$

, where $|0\rangle_x$ is free vacuum. On the other hand, in the first quantization theory [5],

$$G_0^{(1)}[1,2] = \int_{x_1}^{x_2} Dx(\sigma) e^{i\int d^2\sigma L_0} = \theta(\tau_1-\tau_2)\delta(p_1^+ - p_2^+)\int D\vec{x}_\perp e^{i\int_{\Sigma_0} d^2\sigma L_0}\delta[\vec{x}_\perp(\sigma,\tau_1)-\vec{x}_{\perp 1}(\sigma)]\cdot$$

$$\cdot\delta[\vec{x}_\perp(\sigma,\tau_2)-\vec{x}_{\perp 2}(\sigma)] \tag{1}$$

$$L_0 = \frac{1}{2}(\dot{x}^2 - x'^2)$$

, where $\Sigma_0$ is the rectangular region in the world-sheet (Fig. 1).

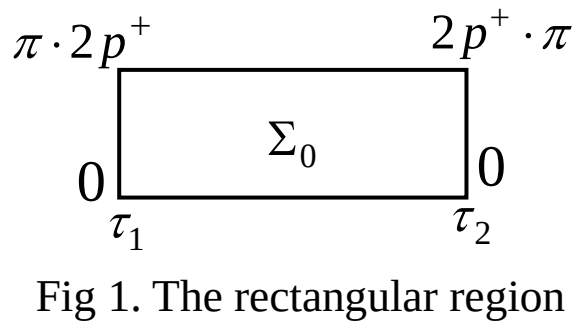

Fig 1. The rectangular region $\Sigma_0$ in the world-sheet

Second, it is possible that factor $\delta[x_1, x_2, x_3]$, which represents splitting and joining of string) in interaction Lagrangian of string field theory can be expressed as functional in $\sigma-\tau$ worldsheet [5]. In case of string's split,

* Corresponding Author
*Email address*: ryongnam13@yahoo.com (Tae-Song Kim)

$$\delta[x_1,x_2,x_3]=\delta[\vec{x}_{\perp 3}(\sigma_3)-\vec{x}_{\perp 1}(\sigma_1)\theta_1-\vec{x}_{\perp 2}(\sigma_2)\theta_2]=$$

$$=\lim_{\substack{\tau_1,\tau_2\to\tau_0+0\\ \tau_3\to\tau_0-0}}\theta(\tau_1-\tau_0)\theta(\tau_2-\tau_0)\theta(\tau_0-\tau_3)\int D\vec{x}_\perp(\sigma,t)e^{i\int_{\Sigma_I}d^2\sigma L_0(\vec{x}_\perp(\sigma,\tau))}\cdot \tag{2}$$

$$\cdot\delta[\vec{x}_\perp(\sigma_1,\tau_1)-\vec{x}_{\perp 1}(\sigma_1)]\delta[\vec{x}_\perp(\sigma_2,\tau_2)-\vec{x}_{\perp 2}(\sigma_2)]\delta[\vec{x}_\perp(\sigma_3,\tau_3)-\vec{x}_{\perp 3}(\sigma_3)]$$

and in case that two strings are joining into one, in equation (2) $\tau_1,\tau_2\to\tau_0-0$ and $\tau_3\to\tau_0+0$, correspondingly $\theta(\tau_0-\tau_1)\theta(\tau_0-\tau_2)\theta(\tau_3-\tau_0)$ is multiplied and integral region in world sheet is turned into $\Sigma_1^*$ (Fig 2).

Equation (2) was proved by using green function in region in reference [5].

The relations similar to equation (2) are maintained in covariant field theory of string. Our goals are first to suggest the general proof method without using the concrete form of green function obtained in light-cone gauge, and second to calculate green function by means of light-cone perturbation series with generating functional represented in light cone diagrams.

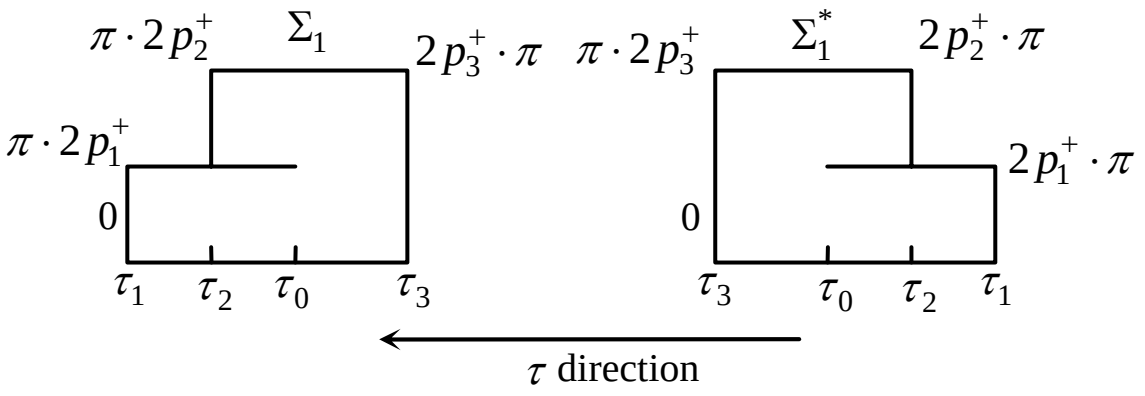

Fig 2. Integral Regions $\Sigma_1$ and $\Sigma_1^*$ in world sheet

## 2. Map of space-time diagram to light-cone diagram

Let's consider the following equation (From now we'll express $\vec{x}_\perp$ as $\vec{x}$ ).

$$\Gamma[x_1,x_2,x_3]\equiv\int\prod_{i=1}^{3}dp_i'^+D\vec{x}_i'(\sigma_i)G_0([\vec{x}_i],p_i^+,\tau_i\,|[\vec{x}_i'],p_i'^+,\tau_0)$$
$$\prod_{\sigma_3=0}^{2\pi p_3^+}\delta(\vec{x}_3'(\sigma_3)-\vec{x}_1'(\sigma_1)\theta_1-\vec{x}_2'(\sigma_2)\theta_2) \tag{3}$$

Integrating this equation with respect to $\vec{x}_3'$ and substituting the definition (1) of $G_0$, we obtain the following:

$$\Gamma[x_1,x_2,x_3]=\int D\vec{x}'(\sigma_1)D\vec{x}_2'(\sigma_2)\prod_{i=1}^{3}dp_i^{+}{}'\delta(p_i'^+-p_i^+)\prod_{j=1}^{2}\theta(\tau_j-\tau_0)\theta(\tau_0-\tau_3)\cdot$$
$$\cdot\int D\vec{x}_1(\sigma_1,\tau)\exp(i\int_{\Sigma_1}d\sigma_1d\tau\,L_0)\delta[\vec{x}_1(\sigma_1,\tau_1)-\vec{x}_1(\sigma_1)]\delta[\vec{x}_1(\sigma_1,\tau_0)-\vec{x}_1'(\sigma_1)]\cdot$$
$$\cdot\int D\vec{x}_2(\sigma_2,\tau)\exp(i\int_{\Sigma_2}d\sigma_2d\tau\,L_0)\delta[\vec{x}_2(\sigma_2,\tau_2)-\vec{x}_2(\sigma_2)]\delta[\vec{x}_2(\sigma_2,\tau_0)-\vec{x}_2'(\sigma_2)]\cdot \tag{4}$$
$$\cdot\int D\vec{x}_3(\sigma_3,\tau)\exp(i\int_{\Sigma_3}d\sigma_3d\tau\,L_0)\delta[\vec{x}_3(\sigma_3,\tau_3)-\vec{x}_3(\sigma_3)]\delta[\vec{x}_3(\sigma_3,\tau_0)-\vec{x}_3'(\sigma_3)]\cdot$$
$$\cdot\delta[\vec{x}_3(\sigma_3,\tau_0)-\vec{x}_1'(\sigma_1)\theta_1-\vec{x}_2'(\sigma_2)\theta_2]$$

Then integrating with respect to $Dx_1'(\sigma_1)$ and $Dx_2'(\sigma_2)$, we obtain:

$$\Gamma[x_1,x_2,x_3]=\theta(\tau_1-\tau_0)\theta(\tau_2-\tau_0)\theta(\tau_0-\tau_3)\int D\vec{x}_1(\sigma_1,\tau)\exp(i\int_{\Sigma_1}d\sigma_1d\tau\,L_0)\cdot$$
$$\cdot\int D\vec{x}_2(\sigma_2,\tau)\exp(i\int_{\Sigma_2}d\sigma_2d\tau\,L_0)\int D\vec{x}_3(\sigma_3,\tau)\exp(i\int_{\Sigma_3}d\sigma_3d\tau\,L_0)\cdot \tag{5}$$
$$\cdot\delta[\vec{x}_1(\sigma_1,\tau_1)-\vec{x}_1(\sigma_1)]\delta[\vec{x}_2(\sigma_2,\tau_2)-\vec{x}_2(\sigma_2)]\cdot$$
$$\cdot\delta[\vec{x}_3(\sigma_3,\tau_0)-\vec{x}_1(\sigma_1,\tau_0)\theta_1-\vec{x}_2(\sigma_2,\tau_0)\theta_2]$$

Now, let's pay attention to the following relation.

$$D\vec{x}_j(\sigma_j,\tau)=\prod_{\tau\in[\tau_j,\tau_0]}\prod_{\sigma_j=0}^{2\pi p_j^+}d\vec{x}_j(\sigma_j,\tau),\quad j=1,2 \tag{6}$$

$$D\vec{x}_3(\sigma_3,\tau)=\prod_{\tau\in[\tau_0,\tau_3]}\prod_{\sigma_3=0}^{2\pi p_3^+}d\vec{x}_3(\sigma_3,\tau)=\prod_{\tau\in(\tau_0,\tau_3]}\prod_{\sigma_3=0}^{2\pi p_3^+}d\vec{x}_3(\sigma_3,\tau)\prod_{\sigma_3=0}^{2\pi p_3^+}d\vec{x}_3(\sigma_3,\tau_0) \quad (7)$$

If equation (5) is integrated with respect to $\prod_{\sigma_3} d\vec{x}_3(\sigma_3,\tau_0)$ considering equation (7), $\delta$-function in equation (5) is removed.

Considering this fact, integral measure remaining in equation (7) is as follows:

$$\prod_{j=1}^{2} D\vec{x}_j(\sigma_j,\tau)D\vec{x}_3(\sigma_3,\tau)=\prod_{\sigma,\tau\in\Sigma_I} d\vec{x}(\sigma,\tau) \quad (8)$$

On one hand, the result integrating equation (5) as above is as follows:

$$\exp\{i(\int_{\Sigma_1}+\int_{\Sigma_2}+\int_{\Sigma_3})\}=\exp(i\int_{\Sigma_I}) \quad (9)$$

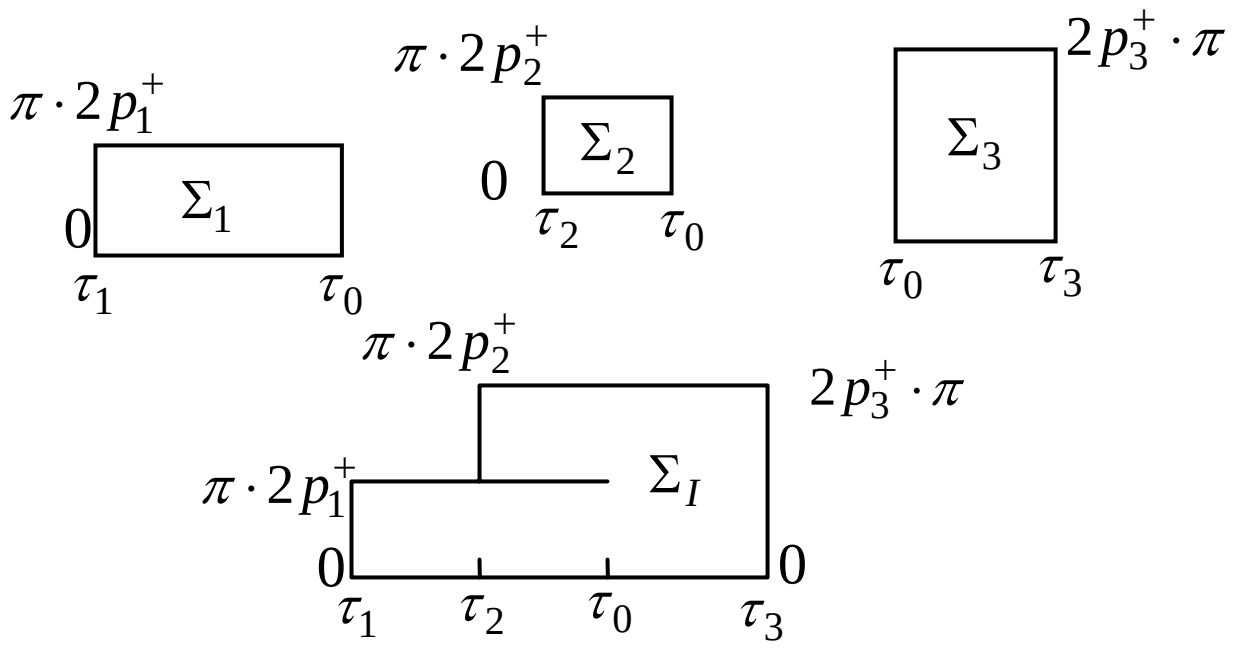

Fig. 3. Integral Regions in World-sheet ($\Sigma_1$, $\Sigma_2$, $\Sigma_3$, $\Sigma_I$)

, where $\Sigma_1$, $\Sigma_2$, $\Sigma_3$, $\Sigma_I$ are shown in Fig 3. Therefore,

$$\Gamma[x_1,x_2,x_3]=\theta(\tau_1-\tau_0)\theta(\tau_2-\tau_0)\theta(\tau_0-\tau_3)\int D\vec{x}(\sigma,\tau)\exp(i\int_{\Sigma_I}d^2\sigma L_0)$$
$$\cdot\delta[\vec{x}(\sigma_1,\tau_1)-\vec{x}_1(\sigma_1)]\delta[\vec{x}(\sigma_2,\tau_2)-\vec{x}_2(\sigma_2)]\delta[\vec{x}'(\sigma_3,\tau_3)-\vec{x}_3(\sigma_3)] \quad (10)$$

On the other hand in equation (3) considering the property of Green function

$$G_0(\tau,p_i^+,[\vec{x}_i],\tau',p_i^{+\prime},[\vec{x}_i'])\Big|_{\tau\to\tau'}=\delta(p_i^+-p_i^{+\prime})\delta[\vec{x}_i-\vec{x}_i'] \quad (11)$$

, and if $\tau_1\to\tau_0+0$, $\tau_2\to\tau_0+0$ and $\tau_3\to\tau_0-0$, the result is as follows.

$$\lim_{\tau_1\to\tau_0+0,\tau_2\to\tau_0+0}\Gamma(x_1,x_2,x_3)=\int\prod_{i=1}^{3}dp_i^{+\prime}\delta(p_i^{+\prime}-p_i^+)\delta[\vec{x}_i'-\vec{x}_i]\delta[\vec{x}_3'-\vec{x}_1'\theta_1-\vec{x}_2'\theta_2]$$
$$=\delta[\vec{x}_3(\sigma_3)-\vec{x}_1(\sigma_1)\theta_1-\vec{x}_2(\sigma_2)\theta_2]= \quad (12)$$
$$=\prod_{\sigma_3=0}^{2\pi p_3^+}\delta(\vec{x}_3(\sigma_3)-\vec{x}_1(\sigma_1)\theta_1-\vec{x}_2(\sigma_2)\theta_2)$$

From equation (10) and equation (12), finally equation (2) is obtained. As shown above, it can be known that the explicit form of Green function was never used and only its definition and properties (11) were used.

So we come to the conclusion that this relation does not belong to only light-cone gauge. This fact is important.

Now, the expression of generating functional [1] producing Feynman's diagrams in space-time can be changed into the generating functional producing light-cone diagram in $\sigma-\tau$ world-sheet.

To do so, we'll put equation (1) of $G_0$ and equation (2) instead of $\delta[x_1,x_2,x_3]$ in $Z_0[J_0,J^*]$.

First, changing $Z_0[J_0,J^*]$,

$$Z_0[J,J^*]=N_0\exp\{i\int d\tau_1 d\tau_2\int_0^\infty dp_1^+dp_2^+\int D\vec{x}_\perp(\sigma)D\vec{y}_\perp(\sigma)J_{p_1^+}(\tau,[\vec{x}_\perp(\sigma)])\cdot$$
$$\theta(\tau_1-\tau_2)\delta(p_1^+-p_2^+)\int D\vec{x}_\perp(\sigma,\tau)\exp(i\int_{\Sigma_0}d^2\sigma L_0)\delta[\vec{x}_\perp(\sigma,\tau_1)-\vec{x}(\sigma)]\cdot$$
$$\delta[\vec{x}_\perp(\sigma,\tau_1)-\vec{y}_\perp(\sigma)]J_{p_2^*}(\tau_2,[\vec{y}_\perp(\sigma)])\}= \quad (13)$$
$$=N_0\exp\{i\int d\tau_1 d\tau_2\theta(\tau_1-\tau_2)\int_0^\infty dp^+\int D\vec{x}_\perp(\sigma,\tau)\exp(i\int_{\Sigma_0}d^2\sigma L_0)\cdot$$
$$J^*_{p^+}(\tau_1,[\vec{x}_\perp(\sigma,\tau_1)])J_{p^+}(\tau_2,[\vec{x}_\perp(\sigma,\tau_1)])\}$$

Second, changing $Z[J_0, J^*]$,

$$Z[J,J^*]=\exp\left\{ i\int d\tau_0 \frac{g}{2}\int\prod_{i=1}^{3}\frac{dp_i^+}{\sqrt{2p_i^+}}\int D\vec{x}_{\perp i}(\sigma_i)\delta(p_3^+-p_1^+-p_2^+)\cdot\right.$$

$$\lim_{\substack{\tau_1,\tau_2\to\tau_0+0\\ \tau_3\to\tau_0-0}}\theta(\tau_1-\tau_0)\theta(\tau_2-\tau_0)\theta(\tau_0-\tau_3)\int D\vec{x}_\perp(\sigma,\tau)\exp(i\int_{\Sigma_I}d^2\sigma L_0)\cdot$$

$$\delta[\vec{x}_\perp(\sigma_1,\tau_1)-\vec{x}_{\perp 1}(\sigma_1)]\delta[\vec{x}_\perp(\sigma_2,\tau_2)-\vec{x}_{\perp 2}(\sigma_2)]\delta[\vec{x}_\perp(\sigma_3,\tau_3)-\vec{x}_{\perp 3}(\sigma_3)]$$

$$\left.(\frac{\delta}{i\delta J_{p_1^+}(\tau_1,[\vec{x}_{\perp 1}])}\frac{\delta}{i\delta J_{p_2^+}(\tau_2,[\vec{x}_{\perp 2}])}\frac{\delta}{i\delta J^*_{p_3^+}(\tau_3,[\vec{x}_{\perp 3}])})+h.c\right\}Z_0[J,J^*]=$$

$$=\exp\left\{ i\int d\tau_0\int\prod_{i=1}^{3}\frac{dp_i^+}{\sqrt{2p_i^+}}\delta(p_3^+-p_1^+-p_2^+)\lim_{\tau_1,\tau_2,\tau_3\to\tau_0}\int D\vec{x}_\perp(\sigma,\tau)[\exp(i\int_{\Sigma_I}d^2\sigma L_0)\cdot\right.$$

$$\cdot\frac{\delta}{i\delta J_{p_1^+}[\vec{x}_\perp(\sigma_1,\tau_1)]}\frac{\delta}{i\delta J_{p_2^+}[\vec{x}_\perp(\sigma_2,\tau_2)]}\frac{\delta}{i\delta J^*_{p_3^+}[\vec{x}_\perp(\sigma_3,\tau_3)]}+\exp(i\int_{\Sigma_I^*}d^2\sigma L_0)\cdot$$

$$\left.\cdot\frac{i\delta}{\delta J^*_{p_1^+}[\vec{x}_\perp(\sigma_1,\tau_1)]}\frac{i\delta}{\delta J^*_{p_2^+}[\vec{x}_\perp(\sigma_2,\tau_2)]}\frac{i\delta}{\delta J_{p_3^+}[\vec{x}_\perp(\sigma_3,\tau_3)]}]\right\}Z_0[J,J^*]$$

Equation (14) considering equation (13) is the generating functional of string field we try to obtain. This generating functional produces the light-cone diagrams in world-sheet.

## 3. Green function of string field

Let's consider the problem calculating Green function by generating functional (14) obtained above. Here we'll find out what kind of series the generating functional (14) produces.
In order to make it simple we'll consider the propagating function of only one string.

$$G^{(1)}[\vec{x}_a(\sigma_a,\tau_a)|\vec{x}_b(\sigma_b,\tau_b)]=\frac{\delta}{i\delta J_{p_a^+}[\vec{x}(\sigma_a,\tau_a)]}\frac{\delta}{i\delta J^*_{p_b^+}[\vec{x}(\sigma_b,\tau_b)]}Z[J,J^*]\Big|_{J=J^*=0} \qquad (15)$$

The term with the lowest order in perturbation expansion of equation (15) gives the propagating function and the second-order term does the one-loop diagram. The result is as follows:

$$G^{(2)}[\vec{x}_a(\sigma_a,\tau_a),p_a^+|\vec{x}_b(\sigma_b,\tau_b),p_b^+]=g^2\frac{\delta(p_a^+-p_b^+)}{16p_a^+}\int_{\tau_b}^{\tau_a}d\tau'\int_{\tau_b}^{\tau'}d\tau\int_0^{p_a^+}\frac{dp_1^+}{p_a^+-p_1^+}\cdot$$

$$\int D\vec{x}(\sigma,\tau)\exp(i\int_{\Sigma^{(2)}}d^2\sigma L_0)\delta[\vec{x}(\sigma,\tau_a)-\vec{x}_a(\sigma_a)]\delta[\vec{x}(\sigma,\tau_b)-\vec{x}_b(\sigma_b)] \qquad (16)$$

, where $\Sigma^{(2)}$ is the wedge region (in $\sigma-\tau$ world-sheet) as Fig 4.

If obtaining the higher order terms like this, the shape of the terms is kept in the main (the multiplicity of the integral with respect of $\tau$ and $p^+$ becomes 2n about the *n*-th order term) and the wedge shape where functional integral in world-sheet is performed is just changed. Expressing the Green function (of one string) in schematic form, it is:

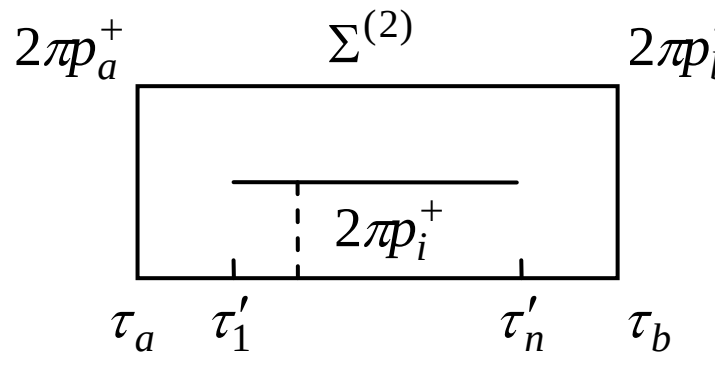

Fig. 4. Wedge Region $\Sigma^{(2)}$

$$G^{(1)}([\vec{x}_a],\tau_a,p_a^+|[\vec{x}_b],\tau_b,p_b^+)=\delta(p_a^+-p_b^+)\sum_{2n=0,2,4,...}(\frac{g}{2})^{2n}$$

$$\int_{\tau_b}^{\tau_a}d\tau_1'\int_{\tau_b}^{\tau_1'}d\tau_1\cdots\int_{\tau_b}^{\tau_a}d\tau_n'\int_{\tau_b}^{\tau_n'}d\tau_n\prod_{i=1}^{n}\int_0^{p_a^+}dp_i^+M(p_1^+,...,p_n^+)\int D\vec{x}(\sigma,\tau) \qquad (17)$$

$$\exp(i\int_{\{\Sigma^{(n)}\}}d^2\sigma L_0)\delta[\vec{x}(\sigma,\tau_a)-\vec{x}_a(\sigma)]\delta[\vec{x}(\sigma,\tau_b)-\vec{x}_b(\sigma)]$$

, where $\{\Sigma^{(n)}\}$ means the sum over the regions of the world-sheet with all possible $n$ split lines. For example, the wedge region possible up to 4th is as follows:

$G^{(1)} =$ 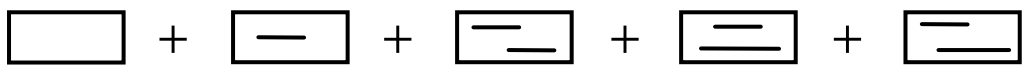 $+ \cdots$ (18)

This equation (18) expresses the light-cone diagram.

Like this in order to obtain the Green function of string field, we have to calculate the functional integral on every kind of the wedge regions considering corresponding boundary conditions on the boundaries and this has to be calculated with conformal mapping [4].

Every region in the equation (18) means the term of perturbation theory and the equation about it is equal to the equation (17).

## Conclusions

We proved the primary relation appearing in transforming the Feynman's diagrams in space-time into the light-cone diagrams in the world-sheet in general form.

By using this, we obtained the generating functional producing the light-cone diagrams and by this equation decided the general form of perturbation series of Green function.

The result gotten in the paper was obtained without especially using the peculiarity in the light-cone gauge and so we have the possibility to transmit this to a different gauge.